\documentclass[draftclsnofoot, onecolumn, 11pt, two-side]{IEEEtran}

\usepackage{times,amsmath,color,amssymb,graphicx,epsfig,cite,psfrag,enumerate,amsthm,multicol,subfigure}

\usepackage[top=0.8in,bottom=0.8in,left=0.8in,right=0.8in]{geometry}

\theoremstyle{remark}
\newtheorem{Remark}{Remark}
\theoremstyle{plain}
\newtheorem{Theorem}{Theorem}
\newtheorem{Proposition}{Proposition}

\theoremstyle{definition}

\begin{document}


\title{Study of Gaussian Relay Channels with Correlated Noises}
%
%
%

\author{Lili~Zhang, 
        Jinhua~Jiang, 
        Andrea J. Goldsmith,
        and~Shuguang~Cui
}
\maketitle
\begin{abstract}
In this paper, we consider full-duplex and half-duplex Gaussian
relay channels where the noises at the relay and destination are
arbitrarily correlated. We first derive the capacity upper bound and
the achievable rates with three existing schemes: Decode-and-Forward
(DF), Compress-and-Forward (CF), and Amplify-and-Forward (AF). We
present two capacity results under specific noise correlation
coefficients, one being achieved by DF and the other being achieved
by direct link transmission (or a special case of CF). The channel
for the former capacity result is equivalent to the traditional
Gaussian degraded relay channel and the latter corresponds to the
Gaussian reversely-degraded relay channel. For CF and AF schemes, we
show that their achievable rates are strictly decreasing
functions over the negative correlation coefficient. Through
numerical comparisons under different channel settings, we observe
that although DF completely disregards the noise correlation while
the other two can potentially exploit such extra information, none
of the three relay schemes always outperforms the others over
different correlation coefficients. Moreover, the exploitation of
noise correlation by CF and AF accrues more benefit when the
source-relay link is weak. This paper also considers the optimal
power allocation problem under the correlated-noise channel
setting. With individual power constraints at the relay and the source, it is shown
that the relay should use all its available power to maximize the
achievable rates under any correlation coefficient. With a total power
constraint across the source and the relay, the achievable rates are
proved to be concave functions over the power allocation factor for AF and
CF under full-duplex mode, where the closed-form power allocation strategy is
derived.
\end{abstract}
\IEEEpeerreviewmaketitle
\renewcommand{\baselinestretch}{1.36}
\section{Introduction} \label{sec_intro}
The relay channel was first introduced by van der Meulen~\cite{van_der},
in which the source communicates with the destination via the help
of a relay node. It was then thoroughly studied in~\cite{cover},
where a max-flow min-cut upper bound and three achievable rates have
been developed for the channel. Among the three achievable rates,
two were obtained with the well-known relaying strategies:
Decode-and-Forward (DF) and Compress-and-Forward (CF), respectively.
The third achievable rate was obtained with a generalized coding
strategy that combines the DF and CF strategies, which is known as
the best achievable rate for the relay channel. Several known achievable rates
meet the upper bound for a few relay channels of special
configurations, leading to the corresponding capacity results. These channels include the degraded and
reversely-degraded relay channels~\cite{cover}, the
semi-deterministic relay channel~\cite{gamal-semi-deterministic},
the phase fading relay channel~\cite{kramer}, the orthogonal relay
channel~\cite{gamal-orthogonal}, and a class of deterministic
channels~\cite{cover-deterministic}. Nevertheless, the capacity of a general relay channel is still an open problem.

In early works on the relay channel, the relay is usually assumed
to be capable of operating in a full-duplex mode, i.e., the relay can
transmit and receive signals at the same time over the same
frequency band. This assumption may not be realistic due to
practical restrictions such as the coupling of transmitted signals
into the receiver path in the RF front-end. As such, more practical
half-duplex relaying has been subsequently studied. In~\cite{gamal},
the half-duplex frequency-division Additive White Gaussian Noise (AWGN) relay channel is
investigated, where the channel is splitted into two frequency bands: One
is utilized by both the source-relay and source-destination links,
and the other is used by the relay-destination link. The
counterpart time-division scheme is presented in~\cite{anders}, where at each slot the relay operates in the receiving mode for the first
$\alpha$ fraction of time and in the transmitting mode for the
remaining $1-\alpha$ fraction. 
The capacity upper bound and the achievable rates for the DF and CF
schemes are derived in both frequency-division and time-division
modes~\cite{gamal,anders}. Another much simpler relaying scheme
under half-duplex mode, Amplify-and-Forward (AF), is discussed
in~\cite{laneman}, where the relay simply forwards a scaled version
of the received signal in the time-division mode with $\alpha =
0.5$.

Most of the existing relay channel work assumes
uncorrelated noises at the relay and the destination. In this work,
we investigate the effect of noise correlation on the Gaussian
relay channel. This situation may arise in practice when the relay
and the destination are interfered with by common random sources.
We consider both the full-duplex and half-duplex
relay channels under the correlated-noise assumption. We extend the
max-flow min-cut upper bound and the DF, CF, and AF coding
strategies to the relay channels with correlated noises. When
applying the DF strategy, the relay decodes the information from the
source, re-encodes and transmits it to the destination. Hence, DF
completely disregards the correlation between the noises at the
relay and the destination. In contrast, when applying the CF
strategy, the relay compresses the received signal (including the
noise) and forwards it to the destination as the side-information,
which may enable us to exploit the noise correlation. Similarly,
when AF is applied, the relay only forwards a scaled version of its
received signal without decoding, and the destination processes the
signals from the source and relay via Maximum-Ratio Combining (MRC),
for which the noise correlation could also be exploited. Although
both CF and AF appear to be more capable in exploiting the noise
correlation, our numerical comparisons under various channel
settings show that none of the coding strategies always outperforms
the others across different correlation coefficients. Beyond the result of achievable rates, we discuss two
capacity-achieving cases with special noise correlations where the achievable rates meet the upper
bound, and they are in fact corresponding to the Gaussian degraded
relay channel and the reversely-degraded one established
in~\cite{cover}. Moreover, it is shown that the noise correlation
always helps with improving the CF and AF performance when $\rho_z \in
[-1,0]$.

Based on the achievable rates for the relay channels with correlated
noises, we also derive optimal power allocation policies under
various constraints. We first consider an individual power
constraint at the relay, and then the total power constraint across both
the source and the relay. Under the individual power constraint at
the relay, it is shown that the relay should always use all its
power to obtain higher achievable rate under any noise correlations.
With the total power constraint, we prove that the power
allocation problems for the CF under full-duplex and AF schemes
are both convex and yield unique solutions, and we provide several closed-form
results for these solutions.

The rest of the paper is organized as follows. The channel model is
discussed in Section \ref{sec_2}. In Section \ref{sec_3}, we provide
the upper bound for the Gaussian relay channel with correlated
noises under both full-duplex and half-duplex modes. The achievable
rates and special capacity-achieving cases are presented in
Section \ref{sec_3}. The power allocation problems are studied in Section \ref{sec_4}. Numerical
results comparing different strategies against the capacity upper bound are presented in
Section \ref{sec_5}. Finally, we conclude the paper in Section
\ref{sec_conclusion}.
\section{Channel Model} \label{sec_2}
\begin{figure}[!t]
\centering
\includegraphics[width=3.5in]{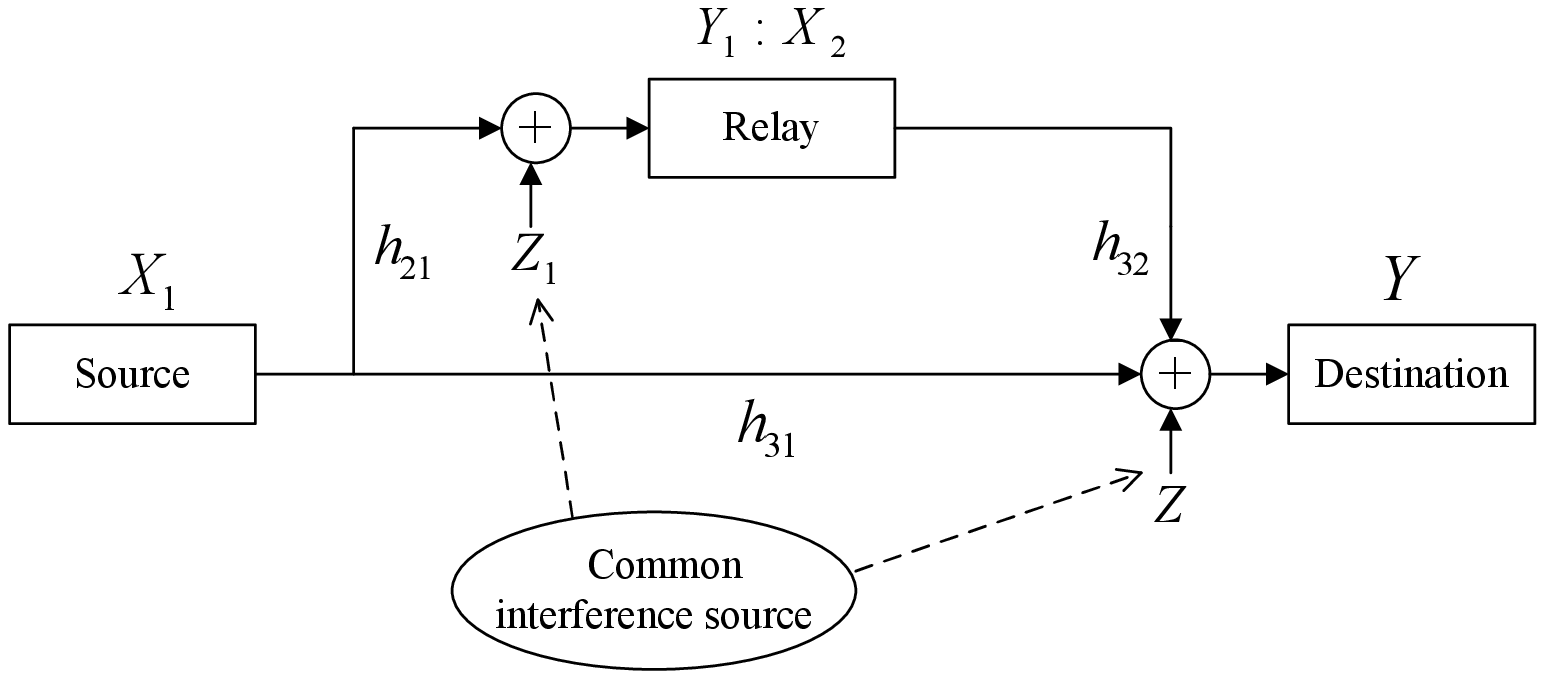}
\caption{The relay channel model with correlated noises at the relay
and the destination.} \label{fig_relay_channel_model}
\end{figure}

Consider the Gaussian relay channel shown in Fig.
\ref{fig_relay_channel_model}, where the source conveys information
to the destination with the help of a relay. The channels
between the respective nodes are assumed to be experiencing static
fading, and the link gains are denoted by $h_{21}$, $h_{31}$, and
$h_{32}$, respectively. The noises that the relay and the
destination suffer from are presumed to be arbitrarily correlated, possibly
due to common interference sources. Under this
correlated-noise channel setting, both full-duplex and half-duplex
operating modes of the relay are investigated.

\subsection{Full-Duplex Mode} \label{subsec_2_1}
In the full-duplex mode, the relay is capable of transmitting and
receiving information at the same time over the same
frequency band. In the $i$th time slot, the channel inputs at the
source and the relay are $X_{1}(i)$ and $X_2(i)$ with power $P_1$
and $P_2$, respectively. The relay experiences an AWGN noise $Z_1(i)$ with variance $N_1$, i.e., $Z_{1}(i) \sim
\mathcal{N}(0,N_{1})$. Similarly, the noise at the destination is denoted as
$Z(i)\sim \mathcal{N}(0, N)$. We further assume that $Z_1$ and $Z$
are arbitrarily correlated, i.e., jointly Gaussian with
correlation coefficient:
\[\rho_z := \frac{\mathbb{E}\{Z_1Z\}}{\sqrt{N_1N}}.\]

Accordingly, the channel under full-duplex mode is fully
described by the following two equations:
\begin{eqnarray*}
Y_{1}(i) &=& h_{21}X_{1}(i)+Z_{1}(i), \\Y(i) &=&
h_{31}X_{1}(i)+h_{32}X_{2}(i)+Z(i),
\end{eqnarray*}
where $Y_{1}(i)$ and $Y(i)$ are the received signals at the relay
and the destination, respectively.
\subsection{Half-Duplex Mode} \label{subsec_2_2}
In practice, it is technically challenging to implement the
relay in the full-duplex mode under which the relay transmits and
receives signals simultaneously within the same frequency band. In
contrast, the frequency division or time-division half-duplex mode
is more practical~\cite{gamal, anders}. In this paper, we consider the time-division half-duplex mode, where at each time slot
the relay listens to the source during the first $\alpha$ fraction of time,
and transmits to the destination in the remaining $1-\alpha$ fraction.

Specifically, in the first $\alpha$ portion of the $i$th time slot, the
transmitted signal at the source is denoted by $X_{1}^{(1)}(i)$ with
an average power constraint $P_{1}^{(1)}$. The received signals at
the relay and the destination are given by:
\begin{eqnarray*}
Y_{1}\left(i\right) &=& h_{21}X_{1}^{(1)}\left(i\right) + Z_{1}\left(i\right),\\
Y^{(1)}\left(i\right) &=& h_{31}X_{1}^{(1)}\left(i\right) +
Z^{(1)}\left(i\right),
\end{eqnarray*}
respectively, where $Z_{1}(i) \sim \mathcal{N}(0,N_{1})$ and
$Z^{(1)}(i) \sim \mathcal{N}(0,N)$ are correlated AWGN noises with
correlation coefficient
\[
\rho_{z} := \frac{\mathbb{E}\{Z_{1}Z^{(1)}\}}{\sqrt{N_{1}N}}.
\]

In the remaining $1-\alpha$ portion, the transmitted signals at the
source and the relay are denoted as $X_{1}^{(2)}(i)$ and $X_{2}(i)$,
with average power constraints $P_{1}^{(2)}$ and $P_{2}$,
respectively. In this phase, the received signal at the destination
is given by:
\begin{eqnarray*}
Y^{(2)}\left(i\right) &=& h_{31}X_{1}^{(2)}\left(i\right) +
h_{32}X_{2}\left(i\right) + Z^{(2)}\left(i\right),
\end{eqnarray*}
where $Z^{(2)}(i) \sim \mathcal{N}(0,N)$ is the AWGN noise at the
destination. With memoryless assumption, we have $Z_1(i)$ and $Z^{(2)}(i)$ independent of each other.

\section{Capacity Upper Bounds and Achievable Rates} \label{sec_3}
In this section, we provide the upper and lower bounds on
the capacity for both full-duplex and half-duplex Gaussian relay
channels with correlated noises. Specifically, max-flow min-cut capacity upper bound is investigated
and various relay strategies (DF, CF and AF) are extended to the respective channel settings to obtain
the corresponding achievable rates. In addition, we characterize the exact capacity of two special
cases, one being achieved by DF and the other achieved by direct
transmission (or a special case of CF). These two cases correspond
to the Gaussian degraded relay channel and the reversely-degraded
relay channel established in~\cite{cover}, respectively.
Furthermore, it is shown that the achievable rates for the CF and AF
schemes are monotonically decreasing functions over negative correlation
coefficient $\rho_z$, which implies that negative noise correlation is always helpful for both the CF and AF
schemes. At the end of this section, we present an alternative channel
model for the correlated-noise channel setting. The alternative relay model has a different source-relay link gain and
independent noises at the relay and the destination, while it has the
same achievable rate as the original correlated-noise
model, when CF or AF scheme is applied.

Note that in this section, all the results are based on a fixed
power assignment: $\{P_{1}, P_{2}\}$ under the full-duplex mode or
$\{P_{1}^{(1)}, P_{1}^{(2)}, P_{2}\}$ under the half-duplex mode; the
case with optimal power allocation will be investigated in Section
\ref{sec_4}. For convenience, we define the following function:
\[ \Gamma \left(x\right) =
\frac{1}{2} \log_{2} \left(1 + x\right).
\]
Also denote the \emph{normalized channel gains} in different links
as:
\begin{equation}
\nonumber \gamma_{21} = \frac{h_{21}^{2}}{N_{1}},\quad \gamma_{32} =
\frac{h_{32}^{2}}{N}, \quad \gamma_{31} = \frac{h_{31}^{2}}{N}.
\end{equation}

\subsection{Max-Flow Min-Cut Upper Bound}\label{subsec_3_1}
\subsubsection{Full-Duplex Mode}
By applying the max-flow min-cut theorem~\cite[Theorem 4]{cover}, we derive an upper bound for the proposed channel model under
the full-duplex mode as follows.
\begin{Proposition}[\textbf{Upper bound with full-duplex}]
For the full-duplex Gaussian relay channel with noise correlation
defined in Section \ref{subsec_2_1}, an upper bound on the capacity
is given by:
\begin{equation}
C^{+}_{\textrm{full}} = \max_{0\leq \rho_x \leq
1}\min\left\{C^{+}_{\textrm{1-full}}(\rho_x),
C^{+}_{\textrm{2-full}}(\rho_x)\right\}, \label{eq_full_upperbound}
\end{equation}
where,
\begin{eqnarray}
\nonumber C^{+}_{\textrm{1-full}}(\rho_x) &=&
\Gamma\left(\gamma_{31}P_1 +
\gamma_{32}P_2+2\rho_x\sqrt{\gamma_{31}P_1\gamma_{32}P_2}\right), \\
\nonumber C^{+}_{\textrm{2-full}}(\rho_x) &=&
\Gamma\left(\frac{P_1(1-\rho_x^2)(\gamma_{21}+\gamma_{31}-
2\rho_z\sqrt{\gamma_{21}\gamma_{31}})} {(1-\rho_z^2)}\right).
\end{eqnarray}
\label{Prop_MFMC_full_upper_bound}
\end{Proposition}
\vspace{-0.3in}
The proof is a trivial extension of that in~\cite[Theorem 4]{cover}, which is skipped here.
\subsubsection{Half-Duplex Mode}
As an extension of the max-flow min-cut upper bound
in~\cite[Proposition 1]{anders}, we have the following upper bound
for the half-duplex mode in the correlated-noise case.
\begin{Proposition}[\textbf{Upper bound with half-duplex}]
For the time-division half-duplex Gaussian relay channel with noise
correlation defined in Section \ref{subsec_2_2}, assuming a fixed
time-division parameter $\alpha$, an upper bound on the channel
capacity is given by:
\begin{equation}
C^{+}_{\textrm{half}} = \max_{0 \leq \rho_{x} \leq 1} \min
\{C_{\textrm{1-half}}^{+}\left(\rho_{x}\right),
C_{\textrm{2-half}}^{+}\left(\rho_{x}\right)\},
\end{equation}\label{eq_half_upperbound}
where
\begin{eqnarray*}
C_{\textrm{1-half}}^{+}\left(\rho_{x}\right)&=& \alpha \Gamma
\left(\gamma_{31}P_1^{(1)}\right) + \left(1-\alpha\right) \Gamma
\bigg(\gamma_{31}P_1^{(2)} + 
\gamma_{32}P_2 + 2\rho_{x}\sqrt{\gamma_{31}P_1^{(2)}\gamma_{32}P_2}\bigg); \nonumber \\
C_{\textrm{2-half}}^{+}\left(\rho_{x}\right)&=& \alpha \Gamma
\left(\frac{\left(\gamma_{21}+\gamma_{31}-2\rho_z\sqrt{\gamma_{21}\gamma_{31}}\right)P_1^{(1)}}{1-\rho_z^2}\right)\nonumber 
+\ \left(1-\alpha\right) \Gamma
\left(\left(1-\rho_x^2\right)\gamma_{31}P_1^{(2)}\right).
\end{eqnarray*}
\label{Prop_MFMC_half_upper_bound}
\end{Proposition}
\vspace{-0.3in}
The proof is a trivial extension of that in~\cite[Proposition 1]{anders}, which is skipped here.

We next study the achievable rates of different coding
strategies, including DF, CF, and AF under both full-duplex and
half-duplex modes.
\subsection{Decode-and-Forward}\label{subsec_3_2}
\subsubsection{Full-Duplex Mode}
In the DF scheme, the relay decodes the message from
the source and re-transmits it to the destination. In fact,
the message is sent to the destination node through cooperation,
which is enabled by the common information at the source and relay nodes.
The following rate is achievable with the DF scheme~\cite{cover}:
\[
R_{\textrm{df}} = \max_{p(x_1,x_2)} \min \{I(X_1;Y_1|X_2),I(X_1,X_2;Y)\}.
\]

Here we evaluate the above rate under the full-duplex channel setting given in
Section \ref{subsec_2_1} as follows.
\begin{Proposition}[\textbf{Achievable rate for DF with full-duplex}]
For the full-duplex Gaussian relay channel with noise correlation
defined in Section \ref{subsec_2_1}, the achievable rate for the
decode-and-forward strategy is given by:
\begin{equation}
R_{\textrm{df-full}} = \max_{0\leq \rho_x \leq
1}\min\left\{R_{\textrm{1-full}}(\rho_x),
R_{\textrm{2-full}}(\rho_x)\right\},\label{eq_df_full_rate}
\end{equation}
where,
\begin{eqnarray}
\nonumber R_{\textrm{1-full}}(\rho_x) &=& \Gamma\left(\gamma_{21}P_1(1-\rho_x^2)\right), \\
\nonumber R_{\textrm{2-full}}(\rho_x) &=& \Gamma\left(\gamma_{31}P_1
+ \gamma_{32}P_2+2\rho_x\sqrt{\gamma_{31}P_1\gamma_{32}P_2}\right).
\end{eqnarray}
\label{Prop_df_full_rate}
\end{Proposition}
\vspace{-0.3in}
\subsubsection{Half-Duplex Mode}
With half-duplex operation, the message is also sent to the destination through
cooperation between the source and the relay. However, the relay only decodes
the received signal from the source in the first phase of each time
slot. In the second phase, the relay re-encodes the received
message, and then transmits it to the destination in cooperation with the source.

The achievable rate for DF under half-duplex mode is given by the
following proposition.
\begin{Proposition}[\textbf{Achievable rate for DF with half-duplex}]
For the time-division half-duplex Gaussian relay
channel with noise correlation defined in Section \ref{subsec_2_2},
assuming a fixed time-division parameter $\alpha$, the achievable
rate for the decode-and-forward strategy is given by:
\begin{equation}
R_{\textrm{df-half}} = \max_{0 \leq \rho_{x} \leq 1} \min
\{R_{\textrm{1-half}}\left(\rho_{x}\right),
R_{\textrm{2-half}}\left(\rho_{x}\right)\},
\end{equation}
where
\begin{eqnarray*}
R_{\textrm{1-half}}\left(\rho_{x}\right)&=&\alpha \Gamma
\left(\gamma_{21}P_1^{(1)}\right)\nonumber 
+ \left(1-\alpha\right) \Gamma
\left((1-\rho_{x}^{2})\gamma_{31}P_1^{(2)}\right);\\
R_{\textrm{2-half}}\left(\rho_{x}\right) &=& \alpha \Gamma
\left(\gamma_{31}P_1^{(1)}\right) + \left(1-\alpha\right) \Gamma
\bigg(\gamma_{31}P_1^{(2)} + 
\gamma_{32}P_2+2\rho_{x}\sqrt{\gamma_{31}P_1^{(2)}\gamma_{32}P_2}\bigg)
\nonumber.
\end{eqnarray*}\label{Prop_df_half_rate}
\end{Proposition}
\vspace{-0.3in}
The proof for Proposition \ref{Prop_df_half_rate} can be found
in~\cite[Propostion 2]{anders}.
\begin{Remark}
We observe that the DF achievable rates under both full-duplex and
half-duplex modes are independent of $\rho_{z}$, and they coincide with
the rates in the uncorrelated-noise case. The reason is that the
relay completely disregards the noise correlation by decoding the
received signal under both full-duplex and half-duplex modes.
\end{Remark}
\subsubsection{Capacity-Achieving Case}
Based on the upper bound and the achievable rates for the DF scheme, here we present the capacity-achieving
case for DF under both full-duplex and half-duplex modes. We show
that the capacity-achieving case for DF is corresponding to the
Gaussian degraded relay channel in~\cite{cover}.
\begin{Theorem}[\textbf{Capacity-achieving case for DF}]
If the noise correlation coefficient $\rho_z$ is equal to
$\sqrt{\gamma_{31}/\gamma_{21}}$
, the capacity of the relay channel defined in Section \ref{sec_2}
is achieved by the DF strategy under both full-duplex and
half-duplex modes. Such a relay channel is equivalent to the Gaussian degraded relay
channel defined in~\cite{cover}, and the capacity under full-duplex
mode is given by:
\begin{equation}
C_{\textrm{1-full}} = \max_{0\leq \rho_x \leq
1}\min\left\{C_{\textrm{11-full}}(\rho_x),C_{\textrm{12-full}}(\rho_x)\right\}.
\end{equation}
where
\begin{eqnarray*}
C_{\textrm{11-full}}\left(\rho_{x}\right)&=&\Gamma\left(\gamma_{21}P_1(1-\rho_x^2)\right);\nonumber\\
C_{\textrm{12-full}}\left(\rho_{x}\right) &=&
\Gamma\left(\gamma_{31}P_1 +
\gamma_{32}P_2+2\rho_x\sqrt{\gamma_{31}P_1\gamma_{32}P_2}\right).
\nonumber
\end{eqnarray*}
The capacity under half-duplex mode is given by:
\begin{equation}
C_{\textrm{1-half}} = \max_{0 \leq \rho_{x} \leq 1} \min
\{C_{\textrm{11-half}}\left(\rho_{x}\right),
C_{\textrm{12-half}}\left(\rho_{x}\right)\},
\end{equation}
where
\begin{eqnarray*}
C_{\textrm{11-half}}\left(\rho_{x}\right)&=&\alpha \Gamma
\left(\gamma_{21}P_1\right) + \left(1-\alpha\right) \Gamma
\left(\left(1-\rho_x^2\right)\gamma_{31}P_1^{(2)}\right);\\
C_{\textrm{12-half}}\left(\rho_{x}\right) &=& \alpha \Gamma
\left(\gamma_{31}P_1^{(1)}\right) +  \left(1-\alpha\right) \Gamma
\bigg(\gamma_{31}P_1^{(2)}+
\gamma_{32}P_2+2\rho_x\sqrt{\gamma_{31}P_1^{(2)}\gamma_{32}P_2}\bigg).
\nonumber
\end{eqnarray*}
\label{theorem_df_capacity}
\end{Theorem}
\vspace{-0.3in}
\begin{proof}
For the full-duplex mode, we observe that the achievable rate of DF
meets the capacity upper bound by substituting $\rho_{z} =
\sqrt{\gamma_{31}/\gamma_{21}}$ into \eqref{eq_full_upperbound} and
\eqref{eq_df_full_rate}.

Now we show that this capacity-achieving case is actually equivalent to the Gaussian degraded relay channel.
In~\cite{cover}, the Gaussian degraded relay channel is defined as:
\begin{eqnarray*}
Y_1 &=& X_1 + Z_1,\\
Y &=& X_1 + X_2 + Z,
\end{eqnarray*}
where $Z = Z_1 + Z_2$, $Z_1$ and $Z_2$ are both zero-mean Gaussian noises and independent of each other.

In our model, after normalizing over respective channel gains, the channel can be described as:
\begin{eqnarray*}
\frac{Y_1}{h_{21}} &=& X_1 + \frac{Z_1}{h_{21}},\\
\frac{Y}{h_{31}} &=& X_1 + \frac{h_{32}}{h_{31}}X_2 + \frac{Z}{h_{31}}.
\end{eqnarray*}

For the relay channel to be degraded, the source-relay link is better than the source-destination link, and $\frac{Z}{h_{31}} = \frac{Z_1}{h_{21}} + Z_2$, where $Z_1$ and $Z_2$ are independent of each other. Hence the
correlation coefficient between $Z$ and $Z_1$ is
\[
\rho_z = \frac{\mathbb{E}\{Z_1Z\}}{\sqrt{N_1N}} = \frac{h_{31}}{h_{21}}\sqrt{\frac{N_1}{N}} = \sqrt{\frac{\gamma_{31}}{\gamma_{21}}}.
\]

The converse can also be established, i.e., given a noise
correlation coefficient $\rho_z = \sqrt{\gamma_{31}/\gamma_{21}}$,
we could always construct a corresponding degraded Gaussian relay
channel.

The capacity under half-duplex mode and the channel equivalence to the
Gaussian degraded relay channel can be derived in a similar way.
\end{proof}
\subsection{Compress-and-Forward}\label{subsec_3_3}
\subsubsection{Full-Duplex Mode}
In the CF scheme, the relay
compresses the received signal with Wyner-Ziv coding~\cite{wyner}, and then forwards the binning index
to the destination. The destination first decodes the binning index
from the relay, and then decodes the compressed signal by viewing
the direct link transmission as the side information. Afterwards, the
destination decodes the original message by utilizing both the
signal from the direct link and the compressed information provided
by the relay. The following achievable rate is established with the CF scheme~\cite{cover}:
\[
R_{\textrm{cf}} = \sup_{p(\cdot)\in\mathcal{P}^{*}} I(X_1;Y,\hat{Y}_1|X_2),
\]
subject to the constraint
\[
I(X_2;Y) \geq I(Y_1;\hat{Y}_1|X_2,Y).
\]
\begin{Proposition}[\textbf{Achievable rate for CF with full-duplex}]
For the full-duplex Gaussian relay channel with noise correlation
defined in Section \ref{subsec_2_1}, the achievable rate for the
compress-and-forward strategy is given by:
\begin{equation}
R_{\textrm{cf-full}}=\Gamma\left(P_1\left(\gamma_{31} +
\frac{(\rho_z\sqrt{\gamma_{31}} - \sqrt{\gamma_{21}})^2}{1-\rho_z^2
+ N_{\textrm{w-full}}/N_1}\right)\right),\label{eq_cf_full_rate}
\end{equation}
where the \textbf{quantization noise power} $N_{\textrm{w-full}}$ is
computed as
\begin{equation}
N_{\textrm{w-full}} = N_1\frac{(1-\rho_z^2) + \gamma_{31}P_1 +
\gamma_{21}P_1 -
2\rho_{z}\sqrt{\gamma_{21}\gamma_{31}}P_1}{\gamma_{32}P_2}.\notag
\end{equation}
\label{Prop_cf_full_rate}
\end{Proposition}
\vspace{-0.3in}
The proof is similar to that in~\cite{cover} and skipped here.
\subsubsection{Half-Duplex Mode}
The relay compresses the
received signal in the first phase with Wyner-Ziv coding~\cite{wyner}, and then transmits it to the
destination in the second phase. We have the following proposition that gives the achievable rate
for CF under half-duplex mode. The proposition is an extended
version of~\cite[Theorem 6]{cover} and~\cite[Proposition 3]{anders};
so the proof is skipped here.
\begin{Proposition}[\textbf{Achievable rate for CF with half-duplex}]
For the time-division half-duplex Gaussian relay channel with noise
correlation defined in Section \ref{subsec_2_2}, assuming a fixed
time-division parameter $\alpha$, the achievable rate for the
compress-and-forward strategy is given by:
\begin{eqnarray}
\nonumber R_{\textrm{cf-half}} &=& \alpha \Gamma
\left(P_1^{(1)}\left(\gamma_{31}+\frac{(\rho_z\sqrt{\gamma_{31}} -
\sqrt{\gamma_{21}})^2}{1-\rho_z^2 + N_{\textrm{w-half}}/N_1}\right)\right)
+\ \left(1-\alpha\right) \Gamma
\left(\gamma_{31}P_1^{(2)}\right),\label{eq_cf_half_rate}
\end{eqnarray}
where the quantization noise power $N_{\textrm{w-half}}$ is computed as 
\begin{equation}\label{cf_nw}
N_{\textrm{w-half}} =
N_{1}\frac{\left(1-\rho_z^2\right)+\gamma_{21}P_1^{(1)}+\gamma_{31}P_1^{(1)}-2\rho_z\sqrt{\gamma_{21}\gamma_{31}}P_1^{(1)}}{\left(1+\gamma_{31}P_1^{(1)}\right)\left(\left(1+\frac{\gamma_{32}P_2}{1+\gamma_{31}P_1^{(2)}}\right)^{\frac{1-\alpha}{\alpha}}-1\right)}.
\end{equation}
\label{Prop_cf_half_rate}
\end{Proposition}
\vspace{-0.3in}
\begin{Remark}
We see that unlike the DF strategy, the CF achievable rates under
both full-duplex and half-duplex modes depend on the noise
correlation coefficient $\rho_z$, which means that the CF scheme has the potential to exploit the noise correlation.
\end{Remark}
\begin{Remark}
It has been shown in some previous works~\cite{cover,gamal} that the
direct transmission from the source to the destination can achieve a
better rate than the DF strategy if the source-relay link is weaker
than the direct source-destination link. The reason is that the DF
strategy requires the relay to completely decode the message from
the source, which creates a bottleneck on the achievable rate when the link from the source to the relay is relatively weak.
However, the CF strategy does not incur such a problem since the
relay only compresses the received signal. Moreover, the CF strategy always incorporates the direct
transmission as a special case: By setting the compressed signal as
a constant, we have the constraint in~\cite[Theorem 6]{cover} always
satisfied, and the CF achievable rate degrades to the capacity of the
point-to-point link between the source and the destination. The
direct transmission rate under full-duplex mode with the Gaussian
channel setting can be easily computed as
\begin{equation}
R_{\textrm{direct-full}}=I(X_1;Y) =
\Gamma\left(\gamma_{31}P_1\right). \label{eq_DL_full_rate}
\end{equation}
Similarly, under half-duplex mode, the direct link transmission rate can be easily derived as:
\begin{equation}
R_{\textrm{direct-half}}=I(X_1^{(1)};Y^{(1)}) + I(X_1^{(2)};Y^{(2)})
= \alpha\Gamma\left(\gamma_{31}P_1^{(1)}\right) +
(1-\alpha)\Gamma\left(\gamma_{31}P_1^{(2)}\right),
\label{eq_DL_half_rate}
\end{equation}
where we have $R_{\textrm{direct-half}} = \Gamma\left(\gamma_{31}P_1\right) = R_{\textrm{direct-full}}$ if $P_1^{(1)} = P_1^{(2)} = P_1$.
\end{Remark}

Regarding the achievable rate for the CF scheme under both
full-duplex and half-duplex modes, we have the following theorem:
\begin{Theorem}[\textbf{Negative correlation always helps CF}]
For the Gaussian relay channel with noise correlation defined in
Section \ref{sec_2}, the achievable rates of CF under both
full-duplex and half-duplex modes are monotonically decreasing
functions over $\rho_z$ when $\rho_z \in [-1,0]$. Therefore,
negative correlation always increases the CF achievable rates
compared to the case of independent noises with $\rho_z = 0$.
\label{theorem_cf_negative}
\end{Theorem}
\begin{proof}
We start with the CF scheme under full-duplex mode. The same
conclusion can be drawn for the half-duplex mode in a similar way.

Since $\Gamma$ is a monotonically increasing function, it suffices
to show that its argument in \eqref{eq_cf_full_rate} is monotonically decreasing over
negative $\rho_z$. Denote the second item inside $\Gamma$ of
\eqref{eq_cf_full_rate} by $G(\rho_z)$ (while the first one is constant over
$\rho_z$):
\begin{eqnarray}
\nonumber G(\rho_z) &=&
\frac{P_1(\rho_z\sqrt{\gamma_{31}} -
\sqrt{\gamma_{21}})^2}{1-\rho_z^2+N_w/N_1}\\
&\triangleq& \frac{G_1\rho_z^2 - G_2\rho_z +
G_3}{-G_4\rho_z^2-G_5\rho_z+G_6}, \ \textrm{with} \  G_i>0, i =
1,...6,\ \textrm{when}\ \rho_z\ \textrm{is negative}.\label{eq_cf_negative}
\end{eqnarray}

The first order derivative of $G(\rho_z)$ over $\rho_z$ is
\begin{equation}\label{eq_cf_g_deri}
G'(\rho_z) = \frac{-(G_1G_5+G_2G_4)\rho_z^2+2(G_1G_6+G_3G_4)\rho_z+G_3G_5-G_2G_6}{(-G_4\rho_z^2-G_5\rho_z+G_6)^2}
\end{equation}
It can be easily seen that when $\rho_z$ is negative, the derivative of $G(\rho_z)$ is negative with $G_3G_5-G_2G_6<0$.
With the above definition of $G_i$'s, we have that
\begin{equation*}
G_3G_5-G_2G_6 =
\frac{-2(1+\gamma_{31}P_1+\gamma_{32}P_2)\sqrt{\gamma_{21}\gamma_{31}}P_1}{\gamma_{32}P_2}
< 0.
\end{equation*}
Therefore, we claim that the achievable rate for CF under full-duplex
mode is strictly decreasing when the noise correlation is negative.
In other words, the negative correlation always helps compared with
the case of independent noises ($\rho_z  = 0$) under full-duplex mode
when CF is applied.

The proof for the case of CF strategy under half-duplex mode is similar and
hence omitted here.
\end{proof}
\begin{Remark}
Intuitively, the reason for the monotonic decrease in the CF rate
over negative $\rho_z$ is the increase in the effective noise
power perceived at the destination, when the negative $\rho_z$
approaches zero. In other words, the received signal-to-noise-ratio
at the destination is monotonically decreasing when the
correlation coefficient $\rho_z$ is negative, which yields the
conclusion in Theorem \ref{theorem_cf_negative}.
\end{Remark}
\begin{Remark}
For the positive noise correlation ($\rho_z>0$), the achievable rate
for the CF scheme is not a monotonic function any more. To some
extent, we could say that compared with the uncorrelated-noise case, the
positive noise correlation hurts the CF scheme for $\rho_z\in
(0,\rho_z')$, where $\rho_z'$ satisfies $G(\rho_z') = G(0)$. However, the
rate at $\rho_z = 1$ coincides with the rate at $\rho_z = -1$, which
means that the positive correlation may also help when $\rho_z
\in [\rho_z',1]$. The detailed proof of the above conclusions is given in Appendix \ref{appendix_1}.
\end{Remark}
\subsubsection{Capacity-Achieving Case}
Here we present another capacity result, where the capacity
is achieved with either CF or direct link transmission. In fact, this case is equivalent to the Gaussian
reversely-degraded relay channel in~\cite{cover}.
\begin{Theorem}[\textbf{Capacity-achieving case for CF}]
If the noise correlation coefficient $\rho_z$ is equal to
$\sqrt{\gamma_{21}/\gamma_{31}}$
, the capacity of the relay channel defined in Section \ref{sec_2}
is achieved by the direct link transmission from the source to the
destination (a special case of CF) under both full-duplex and
half-duplex modes. This case is equivalent to the Gaussian
reversely-degraded relay channel defined in~\cite{cover}. The
capacity under full-duplex mode is given by:
\begin{equation}
C_{\textrm{2-full}} = \Gamma\left(\gamma_{31}P_1\right),
\end{equation}
and the capacity under half-duplex mode is given by:
\begin{equation}
C_{\textrm{2-half}} = \alpha \Gamma
\left(\gamma_{31}P_1^{(1)}\right) + \left(1-\alpha\right) \Gamma
\left(\gamma_{31}P_1^{(2)}\right).
\end{equation}
\label{theorem_cf_capacity}
\end{Theorem}
\vspace{-0.3in}
\begin{proof}
The achievability can be shown by setting the compressed signal in the relay CF operation as
a constant, which means that the rate of CF strategy in this case is
actually the direct link transmission rate under both full-duplex
and half-duplex modes.

For the converse part, the upper bound $C^{+}_{\textrm{full}}$ under full-duplex mode in \eqref{eq_full_upperbound} is further
bounded by:
\begin{eqnarray}
C^{+}_{\textrm{full}} &\leq &\max_{0\leq \rho_x \leq 1} C^{+}_{\textrm{2-full}}\nonumber\\
&=&\Gamma\left(\frac{(1-\rho_x^2)P_1(\gamma_{21}+\gamma_{31}-
2\rho_z\sqrt{\gamma_{21}\gamma_{31}})}
{(1-\rho_z^2)}\right)\nonumber
\\ & = & \Gamma\left(\frac{P_1(\gamma_{21}+\gamma_{31}-
2\rho_z\sqrt{\gamma_{21}\gamma_{31}})} {(1-\rho_z^2)}\right).
\label{eq_full_converse}
\end{eqnarray}
By substituting $\rho_z=\sqrt{\gamma_{21}/\gamma_{31}}$ into
\eqref{eq_full_converse}, we have
\[C^{+}_{\textrm{full}} \leq \Gamma\left(\gamma_{31}P_1\right),\]
which establishes the converse part under the full-duplex mode.

The converse part under half-duplex mode can be proved by
substituting $\rho_{z} = \sqrt{\gamma_{21}/\gamma_{31}}$ into
\eqref{eq_half_upperbound}:
\begin{eqnarray*}
C^{+}_{\textrm{half}} &\leq& \max_{0 \leq \rho_{x} \leq 1} C_{\textrm{2-half}}^{+}\left(\rho_{x}\right) \\
&=& \max_{0 \leq \rho_{x} \leq 1} \left(\alpha \Gamma
\left(\gamma_{31}P_1^{(1)}\right) + \left(1-\alpha\right) \Gamma
\left(\left(1-\rho_x^2\right)\gamma_{31}P_1^{(2)}\right)\right)\\
&=& \alpha \Gamma \left(\gamma_{31}P_1^{(1)}\right) +
\left(1-\alpha\right) \Gamma \left(\gamma_{31}P_1^{(2)}\right).
\end{eqnarray*}

By combining both the achievability and converse parts, we see
that when $\rho_z = \sqrt{\gamma_{21}/\gamma_{31}}$, the capacity is
achieved by the special CF scheme: the direct link transmission.

The capacity-achieving case here is actually equivalent to the
reversely-degraded Gaussian relay channel defined in~\cite{cover}, which is described as:
\begin{eqnarray*}
Y_1 &=& X_1 + Z_1,\\
Y &=& X_1 + X_2 + Z,
\end{eqnarray*}
where $Z_1 = Z + Z_2$, $Z$ and $Z_2$ are both zero-mean Gaussian noises and independent of each other.

In our model, the channel can be described as:
\begin{eqnarray*}
\frac{Y_1}{h_{21}} &=& X_1 + \frac{Z_1}{h_{21}},\\
\frac{Y}{h_{31}} &=& X_1 + \frac{h_{32}}{h_{31}}X_2 + \frac{Z}{h_{31}}.
\end{eqnarray*}

For the relay channel to be reversely-degraded, the source-destination link is better than the source-relay link, and $\frac{Z_1}{h_{21}} = \frac{Z}{h_{31}} + Z_2$, where $Z$ and $Z_2$ are independent of each other. Therefore, the
correlation coefficient between $Z$ and $Z_1$ is
\[
\rho_z = \frac{\mathbb{E}\{Z_1Z\}}{\sqrt{N_1N}} = \frac{h_{21}}{h_{31}}\sqrt{\frac{N}{N_1}} = \sqrt{\frac{\gamma_{21}}{\gamma_{31}}}.
\]

The converse can also be established, i.e., we could always construct a corresponding reversely-degraded Gaussian relay
channel given a noise correlation coefficient $\rho_z = \sqrt{\gamma_{21}/\gamma_{31}}$.
The capacity under half-duplex mode and the channel equivalence to a
reversely-degraded Gaussian relay channel can be derived in a similar way.
\end{proof}
\subsection{Amplify-and-Forward}\label{subsec_3_4}
In the AF scheme (under half-duplex mode), the source transmits the
information to the relay and the destination in the first phase, and
keeps silent in the second phase. The relay forwards a scaled
version of the received signal to the destination without decoding or compressing. The
destination decodes the information at the end of each time slot by
using an Maximum Ratio Combining (MRC) decoder. The time-division
parameter $\alpha$ is fixed to $0.5$ since the length of the
codewords in both phases \emph{must} be the same.

The achievable rate for AF is given by the following proposition,
for which~\cite[Eq.(12)]{laneman} is a special case with independent
noises.
\begin{Proposition}[\textbf{Achievable rate for AF}]
For the time-division half-duplex Gaussian relay channel with noise
correlation defined in Section \ref{subsec_2_2}, the achievable rate
for the amplify-and-forward strategy is given by:
\begin{equation}
R_{\textrm{af}} = \frac{1}{2} \Gamma
\left(\gamma_{31}P_1^{(1)}+\frac{\gamma_{32}P_2\left(\rho_{z}\sqrt{\gamma_{31}}
-
\sqrt{\gamma_{21}}\right)^2P_1^{(1)}}{1+\gamma_{21}P_1^{(1)}+\gamma_{32}P_2\left(1-\rho_{z}^2\right)}\right).
\label{eq_af_rate}
\end{equation}
\label{prop_af_rate}
\end{Proposition}
\vspace{-0.3in}
Regarding the AF achievable rate, we have a similar observation to
that of the CF scheme, which is stated as follows:
\begin{Theorem}[\textbf{Negative correlation always helps AF}]
For the time-division half-duplex Gaussian relay channel with noise
correlation defined in Section \ref{subsec_2_2}, the achievable rate
of AF is a monotonically decreasing function over $\rho_z$ when
$\rho_z \in [-1,0]$, which means that negative correlation always
increases the AF achievable rate compared to the case of
uncorrelated noises with $\rho_z = 0$. \label{theorem_af_negative}
\end{Theorem}
\begin{proof}
By taking the derivative of the item inside the $\Gamma$ function in
\eqref{eq_af_rate} over $\rho_z$, it can be shown that the first
order derivative is always negative when $\rho_z$ is negative (using
the similar proof as in Theorem \ref{theorem_cf_negative}). Combined
with the monotonicity of the $\Gamma$ function, we conclude that
$R_{\textrm{af}}$ is strictly decreasing when $\rho_z$ is between
$-1$ and $0$.
\end{proof}
\begin{Remark}
In fact, the intuitive reasoning of Theorem \ref{theorem_af_negative}
is given as follows. When the signals from the source and the relay add up
together via MRC, the noise power decreases if the noise correlation
coefficient is negative. Therefore, the signal-to-noise ratio at the destination is
increased, which leads to the improvement of the achievable rate compared with the
uncorrelated-noise case.
\end{Remark}
\subsection{Alternative Relay Model for Compress-and-Forward and
Amplify-and-Forward} \label{subsec_3_5}
\begin{figure}[!t]
\centering
\includegraphics[width=3.5in]{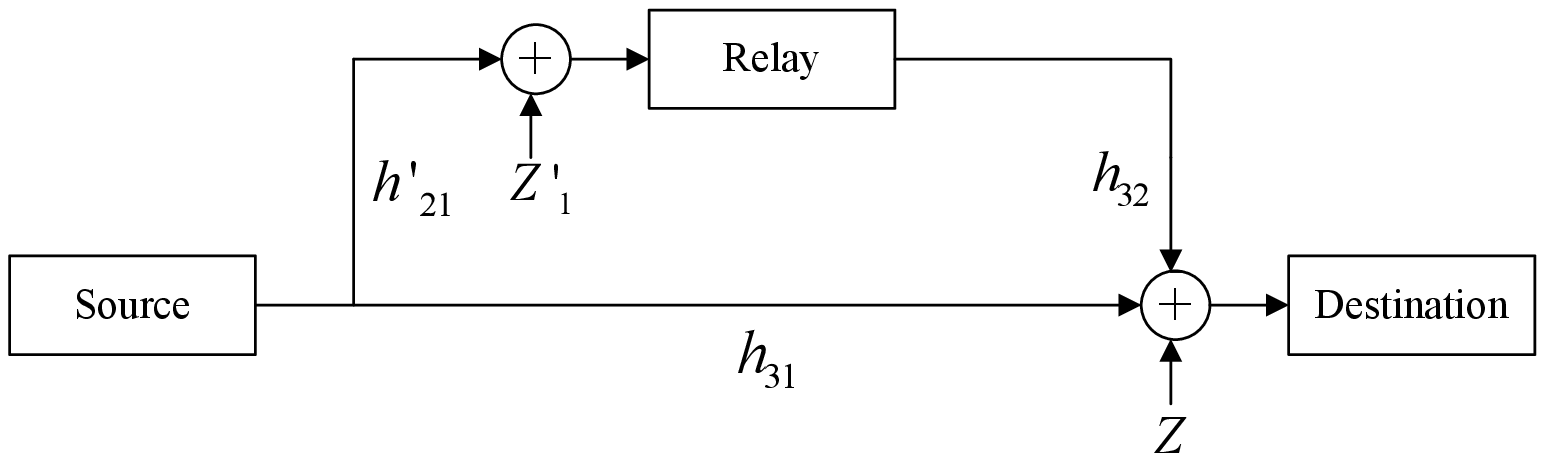}
\caption{The alternative relay channel model for the CF and AF
schemes.} \label{fig_alternative_relay_channel_model}
\end{figure}

From the achievable rates of the CF and AF schemes shown in
\eqref{eq_cf_full_rate}, \eqref{eq_cf_half_rate}, and
\eqref{eq_af_rate}, we observe that they can be achieved by another
alternative relay channel with independent noises and an appropriately defined
source-relay link gain, which is shown
in Fig. \ref{fig_alternative_relay_channel_model}. In particular, the power of
the AWGN noise $Z_1'$ at the relay is $(1-\rho_z^2)N_1$ and the
source-relay channel gain is $h_{21}^{'} = \left|h_{21} -
h_{31}\rho_z\sqrt{\frac{N_1}{N}}\right|$, with other parameters
remaining the same. Under both the full-duplex and half-duplex modes,
the achievable rates of the CF and AF schemes in the alternative relay
model are the same as those for the correlated-noise model discussed previously.

When $\rho_z = 1$ or $\rho_z = -1$, the relay becomes noiseless in
the alternative relay model and the CF achievable rate is
$\Gamma(\gamma_{31}P_1+\gamma_{32}P_2)$ under full-duplex mode.
However, for the AF strategy, the achievable rate at $\rho_z = 1$ is
smaller than the rate at $\rho_z = -1$ since the source-relay link
still affects the achievable rate even with a noiseless relay.

Regarding the special case in Proposition \ref{theorem_cf_capacity},
we find that in the corresponding alternative relay model, the source-relay channel gain
becomes $0$ when $\rho_z = \sqrt{\gamma_{21}/\gamma_{31}}$, such that the
compress-and-forward scheme degrades to the direct link
transmission. Moreover, the conclusions in Theorems
\ref{theorem_cf_negative} and \ref{theorem_af_negative} can be equivalently shown with the alternative relay model, since the normalized
channel gain between the source and the relay is a strictly
decreasing function over negative $\rho_z$.

Such an alternative relay model with independent noises for CF and AF is especially useful when we
derive the optimal power allocation policy for the respective strategies, since we could
extend the results for the traditional relay channel with independent noises, which will be discussed in the next section.

%
\section{Optimal Power Allocation}\label{sec_4}
In this section, we investigate the optimal power allocation at the
source and the relay to maximize the achievable rates of CF and AF
under the correlated-noise channel settings. Two different types of power
constraints are considered: 1) The relay has an individual
power constraint $P_{2\_t}$ with the source power being fixed as
$P_1$; 2) The total power across the source and the relay is
constrained as $P_t$. With the individual power constraint at the
relay, we have the conclusion that the relay should always utilize all
the power budget $P_{2\_t}$ to maximize the achievable rate instead of using
only part of it at an arbitrary noise correlation coefficient $\rho_z$. With
the total power constraint across the source and the relay, we
observe that the achievable rates for the CF scheme under
full-duplex mode and the AF scheme are concave functions over the
power allocation factor. Hence, there exists a unique optimal point and the
closed-form results are provided for the optimal power allocation.
However, with the total power constraint, the power assignment for
the CF strategy under half-duplex mode is not considered here since
no closed-form can be derived~\cite{anders}. We also skip the
optimal power allocation for the DF scheme, which is independent of the
noise correlation and can be referred to~\cite{anders} for more
details.
\subsection{Individual Power Constraint at the
Relay}\label{subsec_4_1}
With an individual power constraint at the relay, we maximize
the achievable rates for CF and AF. When the noises at the relay and the destination
are correlated, it is not clear whether the relay should forward the signal
plus noise with the maximum power. We address this problem as follows.

The optimization problem can be formulated as:
\begin{eqnarray*}
\max_{P_2} && R_{\textrm{cf-full}}\quad \left(\textrm{or} \quad R_{\textrm{cf-half}},\quad R_{\textrm{af}}\right)\\
\textrm{s.t.} && P_2 \leq P_{2\_t},
\end{eqnarray*}
for the CF scheme under full-duplex, half-duplex, and the AF scheme,
respectively.
\begin{Theorem}
For the Gaussian relay channel with noise correlation defined in
Section \ref{sec_2} and an individual power constraint at the relay,
the relay should always use all the power budget $P_{2\_t}$ to achieve higher
rates for both CF and AF strategies at any $\rho_z \in
[-1,1]$.\label{theorem_individual_power}
\end{Theorem}
\begin{proof}
With the achievable rates for CF and AF in
\eqref{eq_cf_full_rate}, \eqref{eq_cf_half_rate}, and
\eqref{eq_af_rate}, it is not straightforward to check whether the relay
should use all of its power. However, with the alternative relay model
defined in Section \ref{subsec_3_5}, we can easily show that the relay should use all of its power
to maximize the CF or AF achievable rates. We take the CF achievable
rate under full-duplex mode as an example, while the same argument holds
for the CF scheme under half-duplex mode and the AF scheme.

When $\rho_z = 1$ or $-1$, the achievable rate for CF under
full-duplex mode is $R_{\textrm{cf-full}} =
\Gamma(\gamma_{31}P_1+\gamma_{32}P_2)$, which is an monotonically increasing
function over the variable $P_2$. Therefore, the relay should use
all the power budget $P_{2\_t}$ if $\rho_z = 1$ or $-1$.

If the noises are not fully-correlated, i.e., $\rho_z \in (-1,1)$, we
adopt the achievable rate under the alternative relay model defined in
Section \ref{subsec_3_5}:
\begin{equation*}
R_{\textrm{cf-full}} =
\Gamma\left(\gamma_{31}P_1+\frac{\gamma_{21}'P_1\gamma_{32}P_2}{1+\gamma_{21}'P_1+\gamma_{31}P_1+\gamma_{32}P_2}\right),
\end{equation*}
where $\gamma_{21}'$ is defined as the effective normalized channel
gain under the alternative relay model:
\begin{equation}
\gamma_{21}' = \frac{\left(\sqrt{\gamma_{21}} -
\rho_z\sqrt{\gamma_{31}}\right)^{2}}{(1-\rho_z^2)}.\label{eq_gamma_21'}
\end{equation}

The above achievable rate for CF under full-duplex mode is also
an increasing function over the variable $P_2$ for $\rho_z \in
(-1,1)$. The same conclusion can be drawn for the CF rate under
half-duplex mode and the AF rate, in similar ways. Therefore, we
conclude that the relay should use all of its power to maximize the
achievable rates under the correlated-noise relay channel setting.
\end{proof}
\subsection{Total Power Constraint}\label{subsec_4_2}
\subsubsection{Compress-and-Forward under Full-Duplex}\label{subsubsec_4_2_1}
The source transmits its
codeword with power $P_1$, and the relay maps the compressed signal
to a new codeword with power $P_2$. The total power is constrained
to $P_t$, i.e., $P_1+P_2 \leq P_t$, such that the optimization problem
can be formulated as:
\begin{eqnarray}
\max_{P_1,P_2} && R_{\textrm{cf-full}}\\
\nonumber\textrm{s.t.} && P_{1} + P_2 \leq
P_t,\label{eq_cf_power_prob}
\end{eqnarray}
where $R_{\textrm{cf-full}}$ is the achievable rate of CF under
full-duplex mode given in \eqref{eq_cf_full_rate}. Since $\Gamma$ function is monotonically increasing, we can
maximize the item inside the $\Gamma$ function instead of
$R_{\textrm{cf-full}}$ itself. The optimal power allocation policy is
provided in the following theorem.
\begin{Theorem}
With a total power constraint across the source and the relay, the
achievable rate for the CF scheme under full-duplex mode is a
concave function over the power allocation parameters $P_1$ and
$P_2$. To maximize the achievable rate in \eqref{eq_cf_full_rate} for the CF scheme under
full-duplex mode, if the noises are fully-correlated ($\rho_z = 1$ or $-1$), the optimal power allocation policy is :
\begin{eqnarray}
\nonumber P_{1}^{*} &=& \left\{\begin{array}{ll}
\epsilon & \textrm{if}\quad \gamma_{31} < \gamma_{32},\ \epsilon\ \textrm{is an arbitrarily small positive number}\\
P_t & \textrm{otherwise,}
\end{array}\right.\label{eq_cf_power_1}\\
P_{2}^{*} &=& P_t - P_{1}^{*}.
\end{eqnarray}

If the noises are not fully-correlated with $\rho_z \in
(-1,1)$, the optimal power allocation is
\begin{eqnarray}
\nonumber P_{1}^{*} &=& \left\{\begin{array}{ll}
\frac{1}{\gamma_{32} - \gamma_{21}' - \gamma_{31}}\left(1+\gamma_{32}P_t-\sqrt{\frac{\gamma_{32}\gamma_{21}'\left(1+\gamma_{32}P_t\right)\left(1+\gamma_{31}P_t+\gamma_{21}'P_t\right)}{\left(\gamma_{32} - \gamma_{31}\right)\left(\gamma_{21}'+\gamma_{31}\right)}}\right) & \textrm{if}\ \gamma_{21}'(\gamma_{32}-\gamma_{31})P_t>\gamma_{31}(1+\gamma_{31}P_t)\\
P_t & \textrm{otherwise,}
\end{array}\right.\label{eq_cf_power_1}\\
P_{2}^{*} &=& P_t - P_{1}^{*},\label{eq_cf_power_2}
\end{eqnarray}
where $\gamma_{21}'$ is the normalized effective source-relay gain
in the alternative relay model defined in \eqref{eq_gamma_21'}.
\label{theorem_cf_full_power}
\end{Theorem}

The detailed proof is given in Appendix \ref{appendix_2} with the
outline sketched as: If the noises are fully-correlated, the achievable rate of
CF under full-duplex mode is $\Gamma(\gamma_{31}P_1+\gamma_{32}P_2)$. If
$\gamma_{31}<\gamma_{32}$, the source power should be as small as
possible, and correspondingly, the relay power should be as large as
possible due to the total power constraint. Otherwise, when
$\gamma_{31}\geq\gamma_{32}$, the source node should use all the
power $P_t$ and the relay keeps silent to maximize the achievable
rate. If the noises are not fully-correlated, first we claim that the constraint
is active at the optimal point: For any given power $P_1$ at the
source, the relay should always utilize all the available remaining power to
obtain higher rate according to Theorem \ref{theorem_individual_power}.
Therefore, we can substitute $P_2$ by $P_t-P_1$ in the objective
function. By taking the second-order derivative of the objective
function over $P_1$ under the alternative relay model mentioned in Section
\ref{subsec_3_5}, we see that the second-order derivative is
always negative, which implies that the objective function is concave
over the variable $P_1$. Then the optimal solution can be solved by
setting the first-order derivative to zero, which results in a
second-order polynomial equation with respect to $P_1$. Due to
the properties of the derived second-order polynomial equation,
either only one of the roots is in the range $(0,P_t)$, or the boundary
point $P_1 = P_t$ is the optimal point, which is
$P_{1}^{*}$ in \eqref{eq_cf_power_1}. Hence,
the power allocated to the relay is $P_{2}^{*} =
P_t-P_{1}^{*}$, which is given in
\eqref{eq_cf_power_2}.
\subsubsection{Amplify-and-Forward}\label{subsubsec_4_2_2}
The source transmits the message
in the first phase and keeps silent in the second phase. The relay
forwards the amplified signal to the destination in the second phase. Due to the fact
that both the source and the relay only use half of the slot for
transmission, the effective average power levels at the source and the relay are
$P_{1}^{(1)}/2$ and $P_2/2$, respectively. As a result, the total power
constraint can be modified as $P_{1}^{(1)} + P_2 \leq 2P_t$. Accordingly, the power
allocation optimization problem can be cast as:
\begin{eqnarray*}
\max_{P_1^{(1)},P_2} && R_{\textrm{af}}\\
\textrm{s.t.} && P_{1}^{(1)} + P_2 \leq 2P_t,
\end{eqnarray*}
where $R_{\textrm{af}}$ is the achievable rate for the AF scheme given in
\eqref{eq_af_rate}.

Similar to the power allocation problem for CF under full-duplex mode, we
consider the item inside the $\Gamma$ function of $R_{\textrm{af}}$
as the objective function, which is also concave over $P_1^{(1)}$ when applying
the alternative relay model defined in Section \ref{subsec_3_5}.
\begin{Theorem}
With a total power constraint over the source and the relay, the
achievable rate for the AF scheme is a concave function over the
power allocation parameters $P_1^{(1)}$ and $P_2$. To maximize the
achievable rate $R_{\textrm{af}}$ for the AF scheme in
\eqref{eq_af_rate}, if the noises are positively fully-correlated
($\rho_z = 1$), the optimal power allocation policy is:
\begin{eqnarray*}
P_{1}^{(1)*} &=& \left\{\begin{array}{ll}
\frac{1}{\gamma_{21}}\left(\left|\sqrt{\gamma_{31}} - \sqrt{\gamma_{21}}\right|\sqrt{\frac{\gamma_{32}\left(1+2\gamma_{21}P_t\right)}{\gamma_{32}\left(\sqrt{\gamma_{31}} - \sqrt{\gamma_{21}}\right)^2-\gamma_{31}\gamma_{21}}}-1\right)& \textrm{if}\quad  \left(\sqrt{\gamma_{31}} - \sqrt{\gamma_{21}}\right)^2 \geq \frac{\gamma_{31}}{\gamma_{32}}\left(\gamma_{21}+\frac{1}{2P_t}\right)\\
2P_t & \textrm{otherwise,}
\end{array}\right.\\
P_{2}^{*} &=& 2P_t - P_{1}^{(1)*}.
\end{eqnarray*}

If the noises are negatively fully-correlated ($\rho_z = -1$), the
optimal power allocation policy is:
\begin{eqnarray*}
P_{1}^{(1)*} &=& \left\{\begin{array}{ll}
\frac{1}{\gamma_{21}}\left(\left(\sqrt{\gamma_{31}} + \sqrt{\gamma_{21}}\right)\sqrt{\frac{\gamma_{32}\left(1+2\gamma_{21}P_t\right)}{\gamma_{32}\left(\sqrt{\gamma_{31}} + \sqrt{\gamma_{21}}\right)^2-\gamma_{31}\gamma_{21}}}-1\right)& \textrm{if}\quad  \left(\sqrt{\gamma_{31}} + \sqrt{\gamma_{21}}\right)^2 \geq \frac{\gamma_{31}}{\gamma_{32}}\left(\gamma_{21}+\frac{1}{2P_t}\right)\\
2P_t & \textrm{otherwise,}
\end{array}\right.\\
P_{2}^{*} &=& 2P_t - P_{1}^{(1)*}.
\end{eqnarray*}

If the noises are not fully-correlated with $\rho_z \in
(-1,1)$, the optimal power allocation is
\begin{eqnarray}
\nonumber P_{1}^{(1)*} &=& \left\{\begin{array}{ll}
\frac{1}{\gamma_{32} - \gamma_{21}'}\left(1+2\gamma_{32}P_t-\sqrt{\frac{\gamma_{32}\gamma_{21}'\left(1+2\gamma_{32}P_t\right)\left(1+2\gamma_{21}'P_t\right)}{\gamma_{32}\gamma_{21}'-\gamma_{31}\gamma_{21}'+\gamma_{31}\gamma_{32}}}\right)& \textrm{if}\quad  \gamma_{21}' \geq \frac{\gamma_{31}}{\gamma_{32}}\left(\gamma_{21}+\frac{1}{2P_t}\right)\\
2P_t & \textrm{otherwise,}
\end{array}\right.\\
P_{2}^{*} &=& 2P_t - P_{1}^{(1)*}.
\end{eqnarray}
where $\gamma_{21}'$ is the normalized effective source-relay gain
in the alternative relay model defined in \eqref{eq_gamma_21'}.
\label{theorem_af_power}
\end{Theorem}

The proof is similar to that of Theorem
\ref{theorem_cf_full_power}, and hence skipped here.
\section{Numerical Results} \label{sec_5}
In this section, we compare the achievable rates of DF, CF, and AF
against the capacity upper bound under both full-duplex and half-duplex
modes. For the full duplex mode, first we consider the fixed power
allocation scenario and set both $P_{1}$ (average power at the
source) and $P_{2}$ (average power at the relay) to $1$. We also
illustrate the performance of the CF scheme using optimal power
allocation. For the half-duplex mode, we first set $\alpha=0.5$, and
compare the achievable rates against the capacity upper bound under fixed power assignment. Then we
present the numerical results after optimizing the
time-division parameter $\alpha$ under half-duplex mode. Finally we
show the improvement due to the optimal power allocation for the AF
scheme. In both full-duplex and half-duplex modes, we set $N = N_{1}
= 1$.

We will apply the following channel model for all the numerical comparisons: the source, relay, and
destination are aligned on a line. The distance between the source
and the relay is $d$ ($0<d<1$), and the distance between the source
and the destination is $1$. The channel amplitude is inversely
proportional to the distance, which means:
\begin{equation}
h_{21} = \frac{1}{d},\ h_{32} = \frac{1}{1-d},\ h_{31} = 1,\quad d
\in (0,1).
\end{equation}
\subsection{Full-Duplex Mode}\label{subsec_5_1}
\subsubsection{Fixed Power Allocation}\label{subsubsec_5_1_1}
\begin{figure*}[!t]
\begin{center}
{\subfigure[Rates vs. $\rho_z$, $d = 0.4$]{\includegraphics[width=3in]{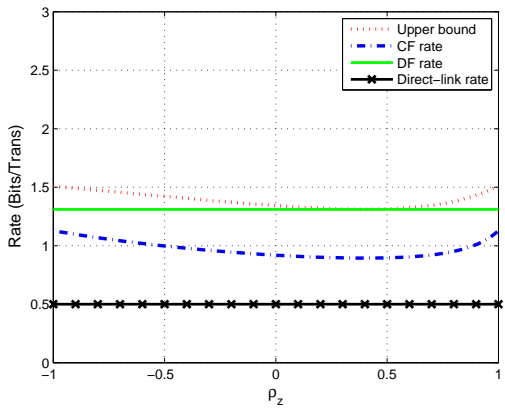}%
\label{fig_rate_vs_rhoz_full_fixed_0.4}}} \hfil
{\subfigure[Rates vs. $\rho_z$, $d = 0.8$]{\includegraphics[width=3in]{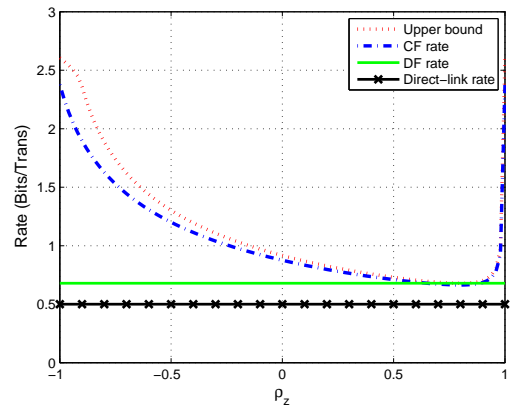}%
\label{fig_rate_vs_rhoz_full_fixed_0.8}}} \caption{Full-duplex: Rates vs.
$\rho_z$, with fixed power allocation.} \label{fig_sim}
\end{center}
\end{figure*}

For the fixed power allocation $\{P_1,P_2\} = \{1,1\}$, we first evaluate
the upper bound and the achievable rates for DF, CF, and direct link
transmission, versus the noise correlation coefficient $\rho_z$.
The cases for different channel parameters ($d= 0.4$ and $d=0.8$)
are shown in Fig. \ref{fig_rate_vs_rhoz_full_fixed_0.4} and Fig.
\ref{fig_rate_vs_rhoz_full_fixed_0.8}, respectively.

From Fig. \ref{fig_rate_vs_rhoz_full_fixed_0.4} and Fig.
\ref{fig_rate_vs_rhoz_full_fixed_0.8}, we see that when the
relay is close to the source ($d = 0.4$), DF always performs the
best, which means that the exploitation of noise correlation is not
necessary since the received SNR at the relay is high enough. However, when
the relay is close to the destination ($d = 0.8$), CF outperforms DF
for most of $\rho_{z} \in [-1,\ 1]$, since the SNR at the relay is
relatively low such that the exploitation of noise correlation
brings more benefits. Furthermore, we observe that the DF rate achieves the capacity upper
bound when $\rho_{z} = \sqrt{\gamma_{31}/\gamma_{21}}$, which
confirms Theorem \ref{theorem_df_capacity}. Meanwhile, Theorem
\ref{theorem_cf_negative} is also verified in these figures, i.e., the achievable rate
for CF under full-duplex mode is strictly decreasing over negative $\rho_z$.

\subsubsection{Optimal Power Allocation for CF}\label{subsubsec_5_1_2}
\begin{figure*}[!t]
\begin{center}
{\subfigure[CF rates vs. $\rho_z$]{\includegraphics[width=3in]{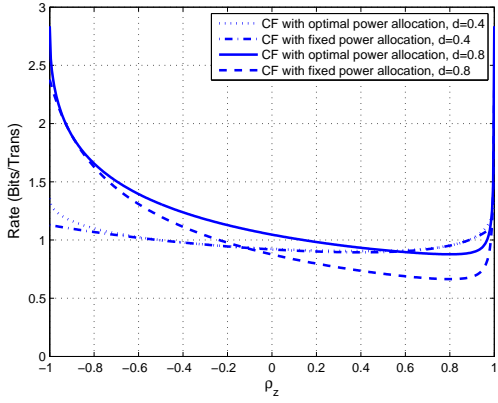}%
\label{fig_rate_vs_rhoz_full_opt}}}\hfil
{\subfigure[Power allocation vs. $\rho_z$]{\includegraphics[width=3in]{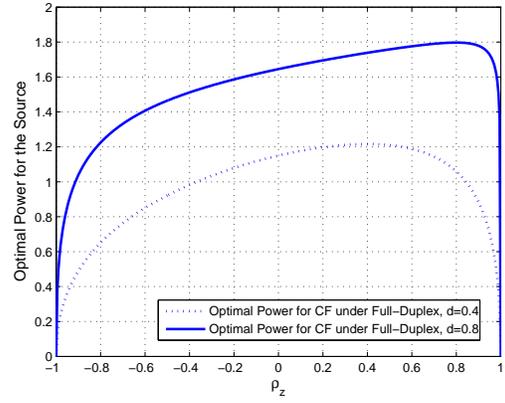}%
\label{fig_power_vs_rhoz_full_opt}}} \caption{Full-duplex: (a) Rates vs. $\rho_z$; (b) Power Allocation vs. $\rho_z$.}\label{fig_CF_rate_vs_rhoz_full}
\end{center}
\end{figure*}

The comparison between the achievable rates for CF under fixed power
allocation and optimal power allocation is shown in Fig. \ref{fig_rate_vs_rhoz_full_opt}, which shows that the optimal
power allocation can bring more profit near fully-correlated
points if the relay is close to the source ($d=0.4$). If the relay
is close to the destination ($d=0.8$), the improvement is more
apparent when the channel is more like a degraded relay channel (near $\rho_z = \sqrt{\gamma_{31}/\gamma_{21}} = 0.8$).

The optimal power allocated to the source are shown in Fig.
\ref{fig_power_vs_rhoz_full_opt} for $d=0.4$ and $d=0.8$, respectively. We see that the source should use more
power if the channel behaves more like degraded, i.e., near $\rho_z=0.4$ when $d=0.4$, and $\rho_z=0.8$ when $d=0.8$, respectively. Moreover, with the same correlation coefficient $\rho_z$,
if the relay is close to the destination ($d=0.8$), the source should use more power
than the case in which the relay is close to the source ($d=0.4$).

%
\subsection{Half-Duplex Mode}\label{subsec_5_2}
\subsubsection{Fixed $\alpha$}\label{subsubsec_5_2_1}
First, the time-division parameter $\alpha$ is set to $0.5$ in the
DF and CF schemes for a fair comparison with AF that has a
requirement of $\alpha = 0.5$. The power assignments are
$\{P_1^{(1)},P_1^{(2)},P_2\} = \{1,1,2\}$ for DF and CF, but
$\{P_1^{(1)},P_1^{(2)},P_2\} = \{2,0,2\}$ for the AF scheme. 

\begin{figure*}[!t]
\begin{center}
{\subfigure[Rates vs. $\rho_z$, $d = 0.4$]{\includegraphics[width=3in]{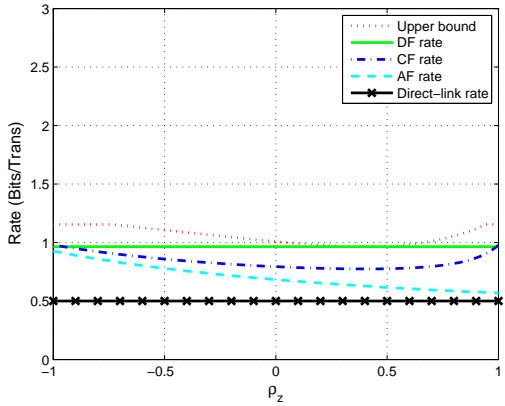}%
\label{fig_rate_vs_rhoz_half_fixed_0.4}}} \hfil
{\subfigure[Rates vs. $\rho_z$, $d = 0.8$]{\includegraphics[width=3in]{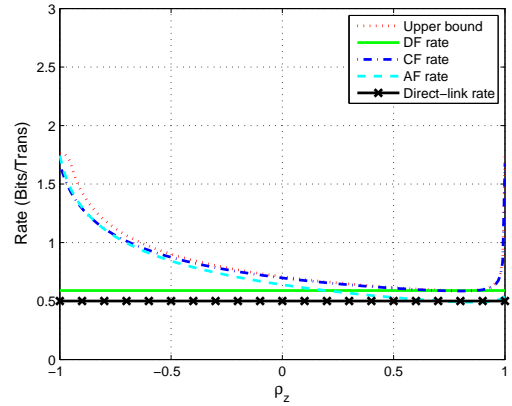}%
\label{fig_rate_vs_rhoz_half_fixed_0.8}}} \caption{Half-duplex: Rates vs.
$\rho_z$, $\alpha = 0.5$, with fixed power allocation.}
\label{fig_sim}
\end{center}
\end{figure*}

The upper bound, the CF rate, the DF rate, the AF rate, and the
achievable rate of direct transmission are plotted in Fig.
\ref{fig_rate_vs_rhoz_half_fixed_0.4} and Fig.
\ref{fig_rate_vs_rhoz_half_fixed_0.8} for different $d$ values. The
analysis of this scenario is similar to the full-duplex case in
Section \ref{subsec_5_1}. In particular, when the relay is close to the
destination, the AF scheme can even beat DF and is close to the
upper bound for all negative $\rho_z$. However, when the
noise correlation coefficient $\rho_z$ is close to $1$, CF
is still the best strategy. When $\rho_z < 0$, Theorems \ref{theorem_cf_negative} and
\ref{theorem_af_negative} are verified by showing that negative
noise correlation helps CF and AF improve the performance. Moreover,
the capacity-achieving case for the DF scheme under half-duplex mode
is confirmed as in Theorem \ref{theorem_df_capacity}.

\subsubsection{Optimization over $\alpha$}\label{subsubsec_5_2_2}
\begin{figure*}[!t]
\begin{center}
{\subfigure[Rates vs. $\rho_z$, $d = 0.4$]{\includegraphics[width=3in]{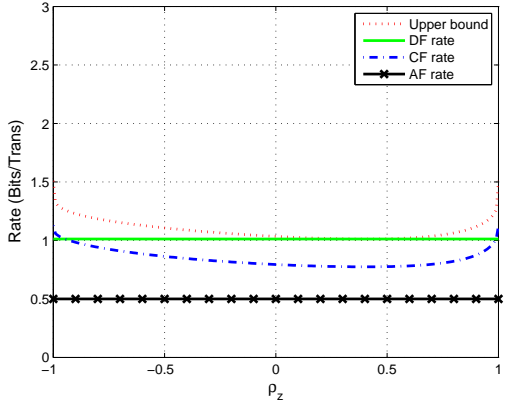}%
\label{fig_rate_vs_rhoz_half_fixed_0.4_alpha_opt}}} \hfil
{\subfigure[Rates vs. $\rho_z$, $d = 0.8$]{\includegraphics[width=3in]{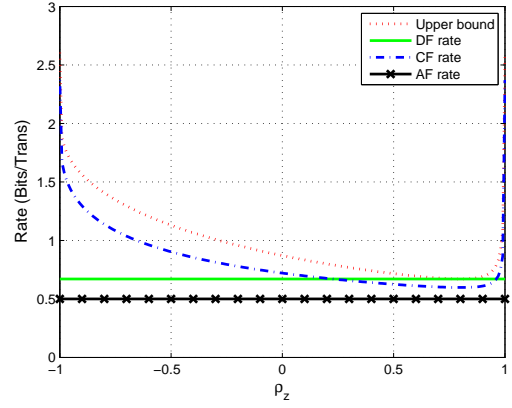}%
\label{fig_rate_vs_rhoz_half_fixed_0.8_alpha_opt}}} \caption{Half-duplex: Rates
vs. $\rho_z$, optimal $\alpha$, with fixed power allocation.} \label{fig_sim}
\end{center}
\end{figure*}

We now consider the improved upper bound and achievable rates via
optimizing $\alpha$ for DF and CF strategies, with the results
shown in Fig. \ref{fig_rate_vs_rhoz_half_fixed_0.4_alpha_opt} and
Fig. \ref{fig_rate_vs_rhoz_half_fixed_0.8_alpha_opt}. The power
allocation is still fixed to $\{P_1^{(1)}, P_1^{(2)}, P_2\} =
\{1,1,2\}$ for both CF and DF.

Compared to the performance of $\alpha = 0.5$ in Fig.
\ref{fig_rate_vs_rhoz_half_fixed_0.4} and Fig.
\ref{fig_rate_vs_rhoz_half_fixed_0.8}, we see that the upper bound
and the achievable rates for DF and CF are increased by optimizing
over $\alpha$. When $d = 0.8$, CF is not dominant over DF any more
for most of $\rho_{z} \in [-1,\ 1]$, which is different from the
performance at $\alpha = 0.5$. The results in Fig. \ref{fig_rate_vs_rhoz_half_fixed_0.4_alpha_opt} and Fig.
\ref{fig_rate_vs_rhoz_half_fixed_0.8_alpha_opt} also show that the
capacity-achieving points for DF are still $\rho_z =
\sqrt{\gamma_{31}/\gamma_{21}}$, which means that Theorem
\ref{theorem_df_capacity} holds
at the optimal $\alpha$. 
\subsubsection{Optimal Power Allocation for AF}\label{subsubsec_5_2_2}
\begin{figure*}[!t]
\begin{center}
{\subfigure[AF rates vs. $\rho_z$]{\includegraphics[width=3in]{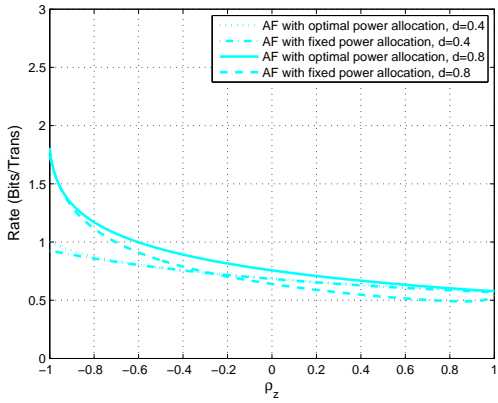}%
\label{fig_rate_vs_rhoz_half_opt}}}\hfil
{\subfigure[Power allocation vs. $\rho_z$]{\includegraphics[width=3in]{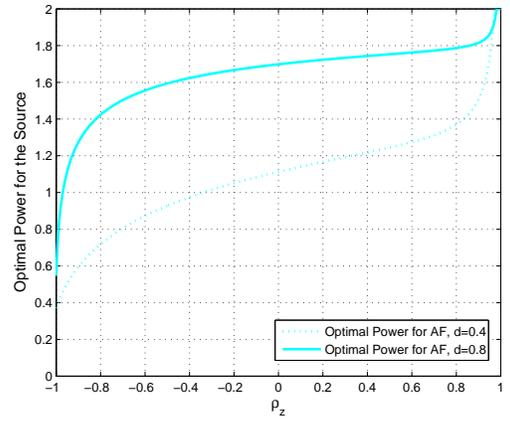}%
\label{fig_power_vs_rhoz_half_opt}}} \caption{Half-duplex: (a) Rates vs. $\rho_z$; (b) Power Allocation vs. $\rho_z$.} \label{fig_AF_rate_vs_rhoz_half}
\end{center}
\end{figure*}

The performance comparisons between AF with optimal power allocation
and AF with fixed power allocation are shown in Fig.
\ref{fig_rate_vs_rhoz_half_opt} for different $d$ values. We can
draw a similar conclusion as in the case of optimal power allocation for CF
under full-duplex mode, except that the optimal power allocation can
no longer bring much benefit when the noise correlation coefficient is
close to $1$ if the relay is close to the source.

The optimal power allocated to the source is shown in Fig.
\ref{fig_power_vs_rhoz_half_opt}. Unlike that of CF in Fig.
\ref{fig_power_vs_rhoz_full_opt}, the source now obtains more and
more power when $\rho_z$ goes to $1$. Basically the reason is that
the relay signal should use less power to forward less noise if the noises are more
positively correlated.

\section{Conclusion} \label{sec_conclusion}
We have obtained upper and lower bounds on the capacity
of Gaussian relay channels with correlated noises at the
relay and the destination. Specifically, the max-flow min-cut upper bound was
first evaluated; then DF, CF, and AF strategies were applied to
derive the respective achievable rates. We also characterized two capacity-achieving cases,
which bear the equivalence to the degraded relay channel and the
reversely-degraded one established in~\cite{cover}, respectively. It was also
shown that the achievable rates of CF and AF are strictly decreasing
functions over negative $\rho_z$, which means that negative
noise correlations always help in CF and AF compared to the
uncorrelated case. 
The optimal power allocation for CF and AF strategies was also investigated. With
an individual power constraint at the relay, the relay should always
use all the power to maximize the
achievable rates under the correlated-noise channel setting. With a
total power constraint over the source and the relay, we proved that the achievable rates for CF
under full-duplex mode and AF are concave functions over the
transmission power and closed-form results were given.
Numerical examples were provided to compare the performance of various coding strategies.

\appendices
\section{}\label{appendix_1}
In this appendix, we prove that the positive correlation may hurt or help the CF achievable rate over two distinct regions.
Recall that the first order derivative of $G(\rho_z)$ in \eqref{eq_cf_negative} is given as
\begin{equation}\label{eq_cf_g_deri_append}
G'(\rho_z) = \frac{-(G_1G_5+G_2G_4)\rho_z^2+2(G_1G_6+G_3G_4)\rho_z+G_3G_5-G_2G_6}{(-G_4\rho_z^2-G_5\rho_z+G_6)^2},
\end{equation}
where the numerator determines the sign of $G'(\rho_z)$. When $\rho_z = 0$, the numerator is reduced to $G_3G_5-G_2G_6 < 0$, which implies that $G'(0) < 0$. When $\rho_z = 1$, the first order derivative of $G(\rho_z)$ is
\begin{eqnarray*}
G'(1) &=& \frac{-G_1G_5-G_2G_4+2G_1G_6+2G_3G_4+G_3G_5-G_2G_6}{(-G_4-G_5+G_6)^2}\\
&=& \frac{2P_1(1+\gamma_{32}P_2+\gamma_{31}P_1)\left(\sqrt{\gamma_{21}}-\sqrt{\gamma_{31}}\right)^2}{\gamma_{32}P_2(-G_4-G_5+G_6)^2} \geq 0.
\end{eqnarray*}
Given that the derivative of $G(\rho_z)$ is negative at $\rho_z=0$ and non-negative at $\rho_z = 1$, and the achievable rate is a continuous function over the
correlation coefficient $\rho_z$, we know that there exists at least one point $\rho_z^* \in [0,1]$ such that $G'(\rho_z^*) = 0$, i.e.,
\[
-(G_1G_5+G_2G_4)\rho_z^{*2}+2(G_1G_6+G_3G_4)\rho_z^*+G_3G_5-G_2G_6 = 0.
\]
By solving the above equation, we have $\rho_z^{*} = \min\{\sqrt{\gamma_{31}/\gamma_{21}}, \sqrt{\gamma_{21}/\gamma_{31}}\}$, which is unique over $[0,1]$. Hence, the CF achievable rate is monotonically decreasing for $\rho_z\in[0,\rho_z^*]$, and monotonically increasing for $\rho_z\in(\rho_z^*,1]$.

As such, when we compare against the uncorrelated-noise case, the noise correlation hurts the CF scheme if $\rho_z \in (0,\rho_z')$, but improves the performance if $\rho_z \in (\rho_z',1]$, where $\rho_z'\in(\rho_z^*,1]$ and $G(\rho_z') = G(0)$. The existence of $\rho_z'$ is unique and can be calculated as
\[
\rho_z' = \frac{G_2G_6-G_3G_5}{G_1G_6+G_3G_4} = \frac{2\sqrt{\gamma_{21}\gamma_{31}}}{\gamma_{21}+\gamma_{31}}.
\]
\section{}\label{appendix_2}
We derive the closed-form solution of the optimal
power allocation for CF under full-duplex mode.
We argue that the constraint in
\eqref{eq_cf_power_prob} must be active at the optimal point, since
for a given $P_1$, the relay should use all the remaining available power to
maximize the achievable rate based on Theorem
\ref{theorem_individual_power}. As such, we substitute $P_2$ by $P_t-P_1$
in the objective function of \eqref{eq_cf_power_prob}.

First assuming the noises are fully-correlated, i.e., $\rho_z = 1$ or
$-1$, the achievable rate of CF under full-duplex mode is $\Gamma(\gamma_{31}P_1+\gamma_{32}P_2)$.
The optimization problem can be cast as:
\begin{eqnarray}
\max_{P_1} && (\gamma_{31}-\gamma_{32})P_1+\gamma_{32}P_t\\
\nonumber\textrm{s.t.} && 0 < P_{1} \leq
P_t.\label{eq_cf_power_prob_new}
\end{eqnarray}
If $\gamma_{31}<\gamma_{32}$, the source power $P_1$ should be as small as
possible (but not zero due to the relay communication setup), and correspondingly, the relay power $P_2$ should be as large as
possible given the total power constraint. Otherwise, when
$\gamma_{31}\geq\gamma_{32}$, the source node should use all the
power $P_t$ and the relay keeps silent to maximize the achievable
rate.

Then we look at the case with $\rho_z \in (-1,1)$. Based on the alternative relay model for the
CF scheme introduced in Section \ref{subsec_3_5}, we optimize the following problem:
\begin{eqnarray}
\max_{P_1,P_2} && \textrm{SNR}_{\textrm{cf-full}} \triangleq \gamma_{31}P_1 + \frac{\gamma_{21}'P_1\gamma_{32}(P_t-P_1)}{1+\gamma_{21}'P_1+\gamma_{31}P_1+\gamma_{32}(P_t-P_1)}\\
\nonumber\textrm{s.t.} && 0 < P_{1} \leq
P_t.\label{eq_cf_power_prob_new}
\end{eqnarray}

For convenience, we define the second item in the objective function
as:
\begin{eqnarray*}
H &\triangleq&
\frac{\gamma_{21}'P_1\gamma_{32}(P_t-P_1)}{1+\gamma_{21}'P_1+\gamma_{31}P_1+\gamma_{32}(P_t-P_1)}\\
&\triangleq& \frac{H_1P_1^2+H_2P_1}{H_3P_1+H_4},\ \textrm{with}\
H_1\leq0,\ H_2\ \textrm{and}\ H_4\geq0
\end{eqnarray*}

The second-order derivative of the objective function is
\begin{equation}
\nabla^2\textrm{SNR}_{\textrm{cf-full}} =
\frac{2H_4(H_1H_4-H_2H_3)}{(H_3P_1+H_4)^3}.
\end{equation}
Since the numerator
\[
H_4 \geq 0,\ H_1H_4-H_2H_3 = -\gamma_{21}'\gamma_{32}-\gamma_{21}'^2\gamma_{32}P_t-\gamma_{21}'\gamma_{31}\gamma_{32}P_t \leq 0,
\]
and the denominator
\[
(H_3P_1+H_4)^3=(\gamma_{21}'P_1+\gamma_{31}P_1-\gamma_{32}P_1+1+\gamma_{32}P_t)^3 > 0,\ \textrm{for}\ 0 < P_{1} \leq P_t,
\]
we see that the second order derivative is non-positive
\begin{equation}
\nabla^2\textrm{SNR}_{\textrm{cf-full}} \leq 0.
\end{equation}

Therefore, the objective function is a concave function over the
variable $P_1$, such that there is a global optimal solution, which is
either the solution of $\nabla\textrm{SNR}_{\textrm{cf-full}} = 0$
or the boundary point $P_t$.

The first-order derivative over $P_1$ can be obtained as
\[
\nabla\textrm{SNR}_{\textrm{cf-full}} = \gamma_{31}+\frac{H_1H_3P_1^2+2H_1H_4P_1+H_2H_4}{(H_3P_1+H_4)^2}.
\]
When $P_1=0$, the first-order derivative $\nabla\textrm{SNR}_{\textrm{cf-full}}=\gamma_{31}+H_2/H_4 > 0$.
If there exists an optimal solution in $(0,P_t)$, the first-order derivative at $P_1=P_t$ must be negative,
\[
\nabla\textrm{SNR}_{\textrm{cf-full}}\big|_{P_1=P_t} = \gamma_{31}- \frac{\gamma_{21}'\gamma_{32}P_t}{1+\gamma_{21}'P_t+\gamma_{31}P_t} < 0,
\]
which implies that $\gamma_{21}'(\gamma_{32}-\gamma_{31})P_t>\gamma_{31}(1+\gamma_{31}P_t)$.
Now with $\nabla\textrm{SNR}_{\textrm{cf-full}} =
0$, we have a second-order polynomial equation over $P_1$,
\[
\left(\gamma_{31}H_3^2+H_1H_3\right)P_1^{*2}+\left(2\gamma_{31}H_3H_4+2H_1H_4\right)P_1^*+\gamma_{31}H_4^2+H_2H_4 = 0,
\]
where only one of the two roots is in the range $(0,P_t)$, and can be written as
the results in \eqref{eq_cf_power_2}.
%

%
%

%

%
%
%




\newpage


%

%

\end{document}